# Implementation of the control and data acquisition system for a small angle neutron scattering spectrometer according to the "Jülich-Munich standard"

H. Kleines[1], M. Drochner[1], H. Lövenich[1], M. Wagener[1], F. Suxdorf[1], F.-J. Kayser[1], K. Zwoll[1],
L. Schätzler[2], J. Heinen[2], M. Heiderich[2], D. Schwahn[2], J. Neuhaus[3], J. Krüger[3]

[1] Zentrallabor für Elektronik, Forschungszentrum Jülich, Germany

[2] Institut für Festkörperforschung, Forschungszentrum Jülich, Germany

[3] FRMII, TU München, Germany

*Abstract*

In Forschungszentrum Jülich the control and data acquisition systems for several neutron spectrometers are being built. Because some of these spectrometers will be commissioned to the new research reactor FRM-II at the technical university of Munich, there was a joint effort with the instrumentation group of the FRM-II to establish the "Jülich-Munich standard", which is basically a collection of tools and devices which are used for the implementation of the spectrometers. This includes: Siemens S7 PLCs for all axis movement issues, PROFIBUS DP for the connection of slow control equipment in the front end, TACO Middleware running on PC-Systems with Linux, python for scripting and Qt for the implementation of GUIs.

The paper describes the implementation the control and data acquisition system of the KWS-1, the first experiment built according to the above standard.

## I. INTRODUCTION

Most neutron scattering experiments at the research reactor FRJ-II in Jülich will be equipped with new electronics and completely new control and data acquisition systems during the next few years. Even a completely new SANS experiment (KWS-3) with a focussing mirror is being developed. Several of these experiments will be transferred to the new high flux neutron source FRM-II in Munich, as soon as it is allowed to start operation. Together with the instrumentation group at the Technical University of Munich, which will operate the FRM-II, the ZEL (Zentrallabor für Elektronik) in Jülich defined a common framework for the control and data acquisition system of all neutron scattering experiments in Jülich and Munich.

Key components of this "Jülich-Munich-Standard" are the consequent use of industrial control technology in the front end and the middleware system TACO, developed by the ESRF for beamline control, on the experiment computers running under Linux [1]. Based on a long term experience in instrumenting Neutron scattering experiments, the ZEL in Jülich concentrated on the development of front end electronics, selection and integration of industrial control technology, development of system software (device drivers,...) and detectors. The instrumentation group in Munich concentrated on application software aspects, especially scripting and extensions of TACO.

In Jülich already two systems, the three axis spectrometer HADAS and the above mentioned KWS-3 (only for first test measurements), are operating according to the new standard, up to the script level. Recently a further SANS (KWS-1) started operation. At KWS-1 also the application software has been implemented according to the new framework, including GUI modules.

In the following sections the framework of the above mentioned standard will be introduced and the implementation of the KWS-1 will be described in detail.

## II. THE "JÜLICH-MUNICH STANDARD"

The "Jülich-Munich standard" is a framework for the selection of technologies and components at each level of the control system. The definition of this framework was motivated by combining development efforts, creation of know how pools and reducing the number of spare parts on the shelf. Up to now Jülich and Munich have exchanged the development results for many hardware components (motor controller, PROFIBUS controller,...) and software modules (device drivers, TACO servers, configuration software,...)

A guiding principle for definition of the framework was to minimize the development efforts and to acquire as much from the market as possible. Slow control in neutron scattering experiments is related to the accurate movement of a diverse range of mechanical parts, to pressure control and temperature control. Because ZEL introduced industrial control equipment already in the 80s to experiment instrumentation, a key component of the framework is the consequent use of industrial technologies like PLCs, fieldbusses or decentral periphery in the front end [2]. Main motivations are:

- low prices induced by mass market,
- inherent robustness
- long term availability and support from manufacturer
- powerful development tools

The following paragraphs list the main components of the framework at each level from bottom to top:

**Field level:** The field level is the lowest level, at which devices that are not freely programmable reside, like motor controllers, PID controllers, analog and digital I/O modules, or measurement equipment. For all industrial type of digital and analog I/Os PROFIBUS DP based decentral periphery is recommend. Siemens ET200S is the preferred one. The only motor controllers supported at the field level are Siemens 1STEP and SMSIPC developed at the University of Göttingen.

**Control level:** The control level resides above the process level. Devices at the control level are freely programmable. They must meet real time requirements and guarantee robust operation in an harsh environment. The only device supported at the control level is the S7-300 PLC family from Siemens, because it dominates the European market. At the control level two additional motor controllers, the Siemens FM357 and FM351, are supported.

**Process communication:** Process communication covers the communication of devices at the field and control level with supervising devices or computers. For lab equipment GPIB and proprietary RS232/RS485 connection are unavoidable. For industrial automation equipment PROFIBUS DP is the recommended choice. It is the dominating fieldbus in Europe and is naturally supported by S7 PLCs and many other devices. A major reason for its success is the technological and functional scalability based on a common core as well as the programming model, which easily maps to PLC operation[3].

**Experiment Computer:** For economical reasons, all experiment computers should be PCs. Linux, being well established in the scientific community, is the only supported operating system. There is no definition of a specific kernel version or distribution. Server computers near the machine should not be conventional desktop PCs but CPCI systems.

**Middleware:** Since the framework aims at an inherently distributed system, software support for the transparent distribution of services between systems is required. For this purpose the TACO has been selected as a middleware system. TACO is a client-server framework developed for beam line control at the ESRF in Grenoble. In a TACO environment each device or hardware module is controlled by a TACO server. The server offers a set of device-specific functions, which can be accessed by TACO clients via an RPC-based mechanism over a TCP/IP network. TACO contains a database for sharing of configuration data between clients and servers.

**Application level:** Flexible operation is a central requirement for neutron spectrometers. For this purpose, spectrometers should be controllable by the script language python. The programming language C++ is recommended for more static applications or modules. Graphical user interfaces (GUI) should be implemented with Qt. Measurement data shall be stored in the NeXus data format.

It is obvious, that this rough framework will be further refined during the implementation of new experiments.

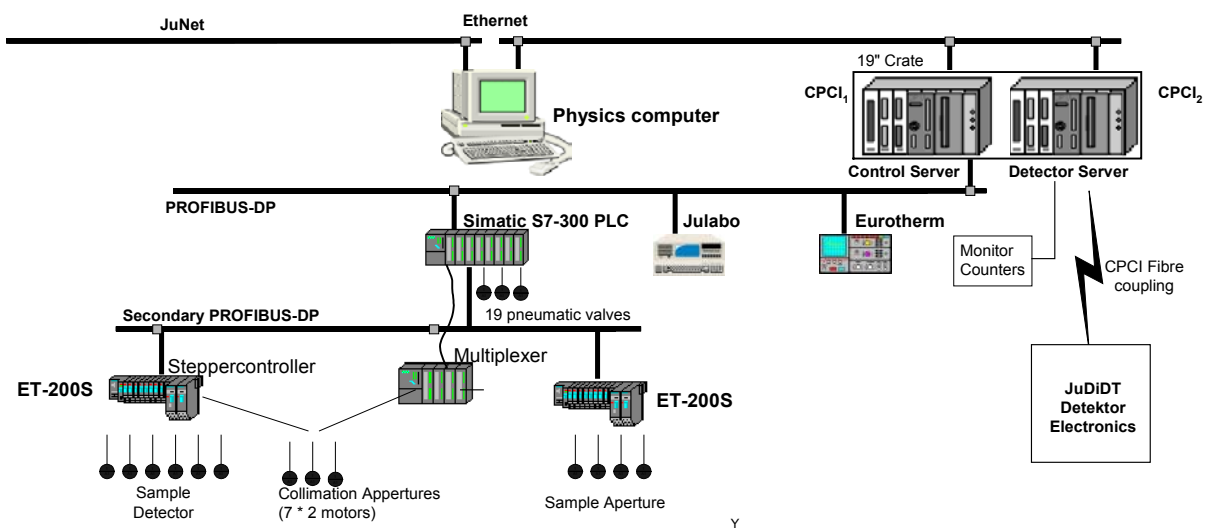

Figure 1: Physical architecture of the control and DAQ system of the KWS 1

# III. THE CONTROL SYSTEM OF THE KWS-1

## 3.1 Structure of the KWS-1

KWS-1 has been designed for a wave number range from $10^{-3}$ to 0.2 Å$^{-1}$ and therefore enables the study of heterogenities of sizes between 10 and 1000 Å. The heterogenities are of different natures such as precipitates and voids in metallic materials, polymers in solution and polymer blends, micelles or microemulsions.

A Dornier velocity selector is used as a monochromator. The collimation can be varied between 1 m and 20 m (by moving 1m neutron guide segments) with 7 variable apertures at fixed positions. Sample environment has an aperture with four axis and is based on a linear sample changer with an additional rotational axis. Ancillary equipment includes several controllers for temperature, pressure and electrical field.

The completely new developed two-dimensional detector is based on the Anger camera principle. A farm of 16 DSPs reconstructs event positions and can be accessed via the CPCI bus [4]. Three-dimensional movement of the detector is possible.

## 3.2 Physical architecture of the control and DAQ system

Accoording to Fig. 1 the control and DAQ system is implemented as a distributed system with an hierarchical architecture. On top of the system resides the physics computer where all application software – GUI-based as well as script-based – is running. It is a Linux-based PC, which is configured as a router between the campus-wide network of FZJ and the separate Ethernet-based experiment network. Via the experiment network the physics computer accesses two server computers. The server computers are CPCI based PCs, which reside in one common CPCI crate. All access to front end systems (detector, monitors, position encoders, motor controllers) is done via these server computers. On the server PCs, which are also Linux-based, TACO servers are running, which access the peripheral devices via dedicated device drivers. The detector and the timer/counter board for reading monitor counts and selector speed are directly accessed via CPCI bus. But all "slow control" periphery is accessed via a subordinate PROFIBUS segment. The most important device on this PROFIBUS is a S7-300 PLC, which controls all mechanical movement. Some of the signals, e.g. the control of the collimation, are directly attached to this PLC. But all motor controllers (exclusively Siemens 1STEP) and SSI interfaces are contained in ET200S decentral periphery systems, which are connected to the PLC via an additional subordinate PROFIBUS segment.

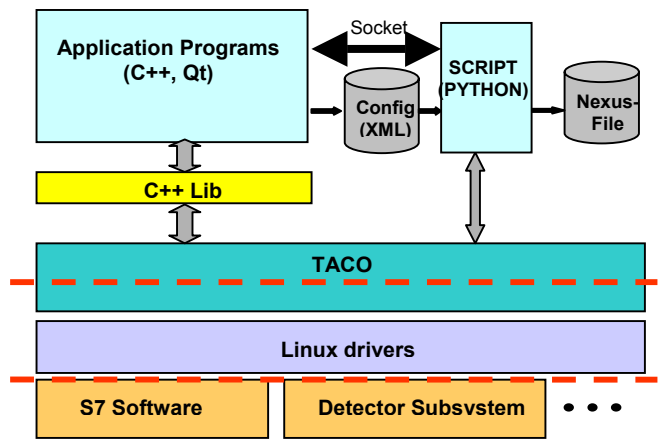

Fig. 2: Software Architecture

## 3.3 Software architecture of the control system

As shown in Fig. 2, the implemented software is distributed between three levels of the system hierarchy. All software below the lower dashed line runs on dedicated front end systems, e.g. PLC software or DSP software on the detector. The software modules shown between the dashed lines are running on the server computers. There basically the TACO servers as well as dedicated device drivers had to be implemented. The TACO middleware is the glue that connects the server computers to the physics computer, where the most complex part of the software (all above the upper dashed line) is running.

The structure of the application software in the physics computer reflects the diverging requirements for spectrometer operation, that is maximum flexibility on one side and maximum degree of productivity and ease of use on the other side. Flexibility is achieved by offering a script interface. Easy usability and high productivity is achieved by a powerful GUI system. Because the KWS-1 is a highly productive machine with lots of external guests, our main focus concentrated on the GUI part, while still offering scripting possibilities.

We developed GUI applications, which allow to define machine setup, samples and complete scans for a given setup, e.g. nested loops over temperature, detector position, sample position. Instead of producing a script out of these definitions, we generate XML files, where all the resulting machine and scan definitions are stored.

We decided to develop one python script, that is well documented, well tested and controls the machine. This script is not static but can be parameterized by reading the above mentioned configuration and definition files in XML format.

Internally, this script offers comfortable, high level interfaces, that can be used by a skilled experimentator to modify this script for specific scans. This allows to maintain modifications to the original script, while still adding new machine configurations via GUI applications.

When the script is running, a further GUI application, the control program, communicates with TACO as well as directly with the script using a private socket-based communication. The control program shows the status of the machine (monitors, axis positions,...) as well as of the script (scan positions, errors). It is able to start the scan, jump to scan positions and to stop the scan.

For service tasks we implemented a GUI application called "manual control", that runs on a notebook with wireless LAN, that allow direct control of all axes.

All GUI applications are based on Qt. They are implemented in C++ with the help of the Qt designer, and use a common library for TACO access, XML parsing and generation, conversion between physical and mechanical units, etc.

## IV. CONCLUSIONS

The implementation of the new control and data acquisition system of the KWS-1 has proven the feasibility of the "Julich-Munich standard". The reuse of software written at the ESRF and at the Technical University of Munich as well as the availability of powerful development tools (Qt designer, Step7, python) have reduced the implementation efforts drastically. A major benefit was the use of professional industrial technologies like PLCs and PROFIBUS.

Still there are some open issues. Because of time restrictions several ad hoc decisions had to be made, which are felt to lack the desired degree of generality, e.g.:
- Application software structure as well as "look and feel" have to be harmonized with Munich developments, as soon as these are more concrete.
- There should be some common high level TACO server for axis movement for all experiments in Julich and in Munich, e.g. for the transformation of machine units to physical units.
- NeXus format has to be implemented and has to be supported by the analysis software.
- The NeXus design team has recently started to formalize the definition of NeXus files using XML. Our XML definitions have to be harmonized with the NeXus XML Meta DTD Format.

Since we have to instrument several spectrometers during the next years, we will consider these aspects in our future developments, and feed all improvements back to the KWS-1. Most of the above items are transparent to the user. But the "look and feel" and the functionality of the application software are a major problem, because there is absolutely no agreement between users.

## V. LITERATURE

[1] A. Götz, et al.: "TACO: An object oriented system for PCs running Linux, WindowsNT, OS-9, LynxOS or VxWorks", Proceedings of the PCaPAC96 Workshop, Hamburg, November 1996

[2] M. Drochner, H. Kleines, P. Wüstner, K. Zwoll, M. Diehl, M. Goldammer, J. Neuhaus: "Application of industrial standard process control equipment in neutron scattering experiments", Proceedings 11[th] IEEE Real Time Conference, Santa Fe, USA, June 1999

[3] H. Kleines, K. Zwoll, M.Drochner, J. Sarkadi: " Integration of Industrial Automation Equipment in Experiment Control Systems via PROFIBUS – Developments and Experiences at Forschungszentrum Jülich", IEEE Transactions on Nuclear Science, Volume 47, Number 2, Part I, April 2000, p. 229-233

[4] G.Kemmerling, et al.:"A New Two-Dimensional Scintillation Detector System for Small-Angle Neutron Scattering Systems", IEEE Transactions on Nuclear Science, Volume 48, Number 4, Part I, August 2001, p. 1114-1117